\documentstyle[11pt,epsfig]{article}
\newcommand{\zp}[3]{Z.\ Phys.\ {\bf C#1} (19#2) #3}
\newcommand{\pl}[3]{Phys.\ Lett.\ {\bf B#1} (19#2) #3}
\newcommand{\np}[3]{Nucl.\ Phys.\ {\bf B#1} (19#2) #3}
\newcommand{\prd}[3]{Phys.\ Rev.\ {\bf D#1} (19#2) #3}

\newcommand{\rd}{{\mathrm{d}}}

\newcommand{\eps}{\epsilon}

\def\simgt{\rlap{\lower 3.5 pt \hbox{$\mathchar \sim$}} \raise 1pt \hbox {$>$}}
\def\simlt{\rlap{\lower 3.5 pt \hbox{$\mathchar \sim$}} \raise 1pt \hbox {$<$}}

\newcommand{\beq}{\begin{equation}}
\newcommand{\eeq}{\end{equation}}
\newcommand{\bea}{\begin{eqnarray}}
\newcommand{\eea}{\end{eqnarray}}

\catcode`\@=11
\def\section{\@startsection{section}{1}{\z@}{3.5ex plus 1ex minus .2ex}
{2.3ex plus .2ex}{\large\bf}}
\def\thesection{\arabic{section}.}

\def\appendix{\setcounter{section}{0}
 \def\thesection{Appendix \Alph{section}:}
 \def\theequation{\Alph{section}.\arabic{equation}}}

\def\@citex[#1]#2{\if@filesw\immediate\write\@auxout{\string\citation{#2}}\fi
  \def\@citea{}\@cite{\@for\@citeb:=#2\do
    {\@citea\def\@citea{,\penalty\@m}\@ifundefined
       {b@\@citeb}{{\bf ?}\@warning
       {Citation `\@citeb' on page \thepage \space undefined}}%
\hbox{\csname b@\@citeb\endcsname}}}{#1}}

\def\citer{\@ifnextchar [{\@tempswatrue\@citexr}{\@tempswafalse\@citexr[]}}
\def\@citexr[#1]#2{\if@filesw\immediate\write\@auxout{\string\citation{#2}}\fi
  \def\@citea{}\@cite{\@for\@citeb:=#2\do
    {\@citea\def\@citea{--\penalty\@m}\@ifundefined
       {b@\@citeb}{{\bf ?}\@warning
       {Citation `\@citeb' on page \thepage \space undefined}}%
\hbox{\csname b@\@citeb\endcsname}}}{#1}}
\relax
\begin{document}
\thispagestyle{empty}
\begin{flushright}
CERN-TH/95-291\\
DESY 95-199\\
November 1995
\end{flushright}
\vskip 1.5cm
\begin{center}{{\bf\Large HEAVY-QUARK CORRELATIONS IN DIRECT \\
\vglue .2cm
PHOTON-PHOTON COLLISIONS}}
\vglue 1.5cm
\begin{sc}
Michael Kr\"{a}mer\\
\vglue 0.3cm
\end{sc}
{\it Deutsches Elektronen-Synchrotron DESY\\
D-22603 Hamburg, Germany}
\vglue 0.3cm
and
\vglue 0.3cm
\begin{sc}
Eric Laenen\\
\vglue 0.3cm
\end{sc}
{\it CERN TH-Division\\
1211-CH, Geneve 23, Switzerland}
\end{center}
\vglue 1.7cm

\begin{abstract}
\noindent
In two-photon collisions at LEP2 and a future $e^+e^-$ linear collider
heavy quarks (mainly charm) will be pair-produced rather copiously. The
production via direct and resolved photons can be distinguished
experimentally via a remnant-jet tag. We study correlations of the
heavy quarks at next-to-leading order in QCD in the direct channel,
which is free from phenomenological parton densities in the photon.
These correlations are therefore directly calculable in perturbative
QCD and provide a stringent test of the production mechanism.

\end{abstract}

\vfill
\begin{flushleft}
CERN-TH/95-291\\
DESY 95-199
\end{flushleft}

\newpage

\setcounter{page}{1}

\noindent

\section{Introduction}
The production of heavy quarks in two-photon collisions has
interesting aspects. Each of the photons can behave as either a
pointlike or a hadronic particle \cite{phhad}. Consequently one
distinguishes in such collisions direct- (both photons are pointlike),
single resolved- (one photon is pointlike, the other hadronlike), and
double resolved (both are hadronlike) production channels. The
resolved channels require the use of parton densities in the photon,
whereas the production via the direct channel is free of such
phenomenological inputs and depends only on the QCD coupling and the
heavy quark mass. The heavy mass provides the hard scale for the
perturbative analysis and ensures that the separation into direct and
resolved production channels is unambiguous even at the
next-to-leading order (NLO) level. Hence production via the direct
channel is directly calculable in perturbative QCD (pQCD) and in
principle the best way for examining the validity of such an analysis
and for confronting the pQCD prediction with experiment.

Two-photon collisions can be investigated at $e^+e^-$ colliders, where
a large number of equivalent photons is generated. Charm quark
production in two-photon collisions has been analysed in many
experiments. One has mainly studied the reaction $e^+e^- \rightarrow
e^+e^- D^{*\pm} X$ with neither outgoing lepton tagged (``no-tag''),
because it proceeds predominantly via the fusion of two equivalent
photons to produce open charm ($\gamma\gamma\rightarrow c\bar c$). The
existence of the $D^{*\pm}$ has been inferred either from direct
reconstruction \cite{dM} or from unfolding the distribution of soft
pions \cite{sp} resulting from its decay. There
have in addition been studies that use soft leptons \cite{sl}
and kaons \cite{k0} to tag charm quarks.

Due to the low experimental acceptance of heavy quark production in
two-photon collisions this reaction has been studied also
theoretically at next-to-leading order in QCD only in the
single-particle inclusive case. Ref.\cite{DKZZ} concentrated on the
no-tag case, and ref.\cite{LRSN} on the case where one of the outgoing
leptons is tagged.  At LEP2 and a future $e^+e^-$ linear collider
(NLC) the higher cms energy and large luminosity will lead to fairly
copious production of charm quark pairs. Thus it will become possible
to measure both heavy quarks and analyse their correlations.  The
study of these correlations constitutes a more comprehensive test of
the theory and is our purpose in this letter. Heavy-quark correlations
have been investigated theoretically also in hadroproduction
\cite{mnr}, photoproduction \cite{fmnr} and electroproduction
\cite{hs}, and experimentally in \cite{correx}. We concentrate here on
the no-tag case and, to eliminate the uncertainties related to the
parton densities in the photon, on the direct channel only. Note that
the TOPAZ collaboration \cite{k0} has recently shown that the direct
channel may be isolated experimentally from the resolved ones by
detecting the photon-remnant jet, present in the resolved channels
only.

The paper is organized as follows: In section 2 we describe our method
of calculation and in section 3 we show heavy quark correlations for
LEP2, and a NLC at a center of mass energy of 500 GeV.  We conclude in
section 4.

\section{Method}
In this section we describe the method we used to calculate the QCD
corrections to the process
\begin{equation}
\gamma(k_1)+\gamma(k_2) \rightarrow Q(p_1)+\overline{Q}(p_2)\, ,
\label{bornproc}
\end{equation}
where $Q$($\overline{Q}$) is a heavy (anti)-quark. We want to have
full exclusive information about the final state. Our method is a
special case of a more general method for performing exclusive higher
order QCD calculations \cite{GGKSL}.

The Born process (\ref{bornproc}) is described by the differential
cross section
\begin{eqnarray}
\rd\sigma^{(0)} & = & \frac{4 \alpha_e^2 e_Q^4 N_c}{s}
\left(\frac{t_1}{u_1}+\frac{u_1}{t_1}+\frac{4 m^2 s}{t_1 u_1}
\left(1-\frac{m^2 s}{t_1 u_1}\right)\right)\nonumber \\
&\times& \frac{\rd^3p_1}{2 \sqrt{|\vec{p_1}|^2+m^2}}\,\frac{\rd^3p_2}
{2 \sqrt{|\vec{p_2}|^2+m^2}}\,\, \delta^{(4)}(k_1+k_2-p_1-p_2)\, .
\label{bornform}
\end{eqnarray}
Here $e_Q$ is the charge of the heavy quark in units of $e$, $N_c=3$
the number of colors, and $m$ the mass of the heavy quark. The
kinematic invariants are defined by
\begin{equation}
s=(k_1+k_2)^2, \;\;\; t_1=(k_1-p_1)^2-m^2, \;\;\; u_1=(k_1-p_2)^2-m^2\, .
\end{equation}
The virtual QCD corrections to the Born process consist of the
interference between the Born amplitude (depicted e.g.\ in Fig.~A1 in
\cite{DKZZ}) and its one-loop corrections. Explicit results have
already been presented in \cite{BKSN} and we will not repeat the
details of the calculation here. We merely note that we regularized
the ultraviolet (UV) and infrared (IR) singularities that occur in the
virtual corrections by working in $d=4-2\epsilon$ dimensions, and
absorbed the UV singularities via mass renormalization in the on-shell
scheme. We are then left with only IR singularities, which appear as
$1/\eps$ poles and factorize into a universal factor multiplying the
Born differential cross section, eq.(\ref{bornform}).

The bremsstrahlung corrections at NLO are due to the radiation of a
gluon from one of the heavy quarks
\begin{equation}
\gamma(k_1)+\gamma(k_2) \rightarrow Q(p_1)+\overline{Q}(p_2)
+ g(k_3)\, .
\label{gluonrad}
\end{equation}
Since our method here is a little different from what was done
previously in the literature, we give a few more details.  Note first
of all that when a gluon is radiated from the heavy quark, no
collinear singularity occurs, because it is shielded by the heavy
quark mass. We divide up the phase space into a ``soft'' region and a
``hard'' region. The soft region is defined by the condition
\begin{equation}
0 \leq s_{13},\,s_{23} \leq s_{\mathrm{min}}
\label{sps}
\end{equation}
where $s_{i3} = 2 p_i\cdot k_3\;(i=1,2)$ and $s_{\mathrm{min}}$ is an
arbitrary cut-off, to be chosen small. The hard region is the
complementary one.

In the hard phase space region, one can work in 4 dimensions and
perform the phase space integrations numerically, allowing for easy
implementation of experimental cuts. As is well known, in the soft
region both the phase space and the matrix element factorize in the
limit of small $s_{\mathrm{min}}$. In both cases, one of the factors
contains the quantum numbers of the gluon, and the other is only
related to the lower order process. As a consequence, one may perform
the integral of the momentum of the gluon in this region analytically
in $d$-dimensions. Specifically one must do the integral
\begin{equation}
4\pi \alpha_s C_F \int {\rm dPS}({\rm soft}) K({\rm soft})\, .
\end{equation}
with the color factor $C_F = (N^2-1)/(2N)$. Here the soft gluon phase
space factor is
\begin{equation}
{\rm dPS}({\rm soft}) =
\frac{(4\pi)^\eps}{16 \pi^2 \Gamma(1-\eps)} \rd s_{13} \rd s_{23}
(s\beta)^{2\eps-1}
[s_{12}s_{13}s_{23}-m^2(s_{13}^2+s_{23}^2)]^{-\eps}
\end{equation}
where $\beta = \sqrt{1-4m^2/s}$ and $s_{12} = 2 p_1\cdot p_2$.  Note
that the expression in square brackets must be positive.  The soft gluon
matrix element factor can easily be found in the eikonal
approximation, and is
\begin{equation}
K({\rm soft}) = 4\left(\frac{s_{12}s_{13}s_{23}-m^2(s_{13}^2+s_{23}^2)}
{s_{13}^2 s_{23}^2}\right)\, .
\end{equation}
Thus, upon combining both factors and integrating over the range
(\ref{sps}), one obtains a universal factor multiplying the
differential Born cross section (\ref{bornform}). This factor contains
$1/\eps$ poles which cancel against those originating in the virtual
corrections. The soft contribution to the fully differential cross
section can then finally be written as
\begin{equation}
 \rd\sigma^{(1)}({\rm soft}) = S_F(s,m^2,s_{\mathrm{min}})
 \,\, \rd\sigma^{(0)}
\end{equation}
where
\begin{eqnarray}
S_F&\!\!=\!\!& \left(\frac{\alpha_s C_F}{\pi}\right)
 \left\{\vphantom{\frac{2m^2}{s}} \right.
 -2\left(1+\left(1-\frac{2m^2}{s}\right)\frac{\ln x}{\beta}
 \right)\left(\ln x -\ln\left(\frac{s}{s_{\mathrm{min}}}\right)-\ln\beta
 \right) \nonumber \\
 &&\hspace*{2cm} -
 2\left(\ln(1-x)+\ln(1+x)-\ln x\right) + 1 -\beta
 \nonumber \\
 &&\hspace*{2cm} -
 \frac{1}{\beta}\left(1-\frac{2m^2}{s}\right)\ln x
 \left(1+2\ln\frac{(1+x)(1-x)}{x}\right) \\
 &&\hspace*{2cm} +
 \frac{1}{2 \beta}\left(1-\frac{2m^2}{s}\right)\left({\rm Li_2}
 \left(1-\frac{1}{x^2}\right)-{\rm Li_2}(1-x^2)\right)
  \nonumber \\
 &&\hspace*{2cm}  \left. +
 \frac{m^2}{s\beta}\ln x \left(\frac{(1+x)(1-x)}{x}+\frac{3s}{2m^2}
 \left(1-\frac{2m^2}{s}\right)\ln x\right)\right\}\,. \nonumber
\end{eqnarray}
Here $x = (1-\beta)/(1+\beta)$ and ${\rm Li_2}(z)$ is the
dilogarithmic function as defined in \cite{lewin}.

Finally, one is left with a two-to-two particle contribution
(consisting of the Born and soft-plus-virtual corrections) and the
two-to-three particle contribution in the hard region.  Each
contribution depends on the theoretical cut-off $s_{\mathrm{min}}$,
but as long as $s_{\mathrm{min}}$ is small enough compared to the
typical scale of our process, the sum does not. This we checked
explicitly.

\section{Results}
Using the method described in the previous section we have constructed
a Monte Carlo program for the reaction $\gamma\gamma \rightarrow
Q\overline{Q}$ for direct photons, including the complete
${\cal{O}}(\alpha_s)$ corrections, which is fully exclusive in all
final state particles. We checked that we could reproduce the results
in \cite{DKZZ} for the total cross section and single particle
transverse momentum ($p_t$) and rapidity ($y$) distributions for the
direct channel. We only present results for charm quark production
because the bottom quark production rate is very much reduced in
two-photon collisions due to charge and phase space suppression.

We first list the default choices we made for various parameters for
producing the results shown in the rest of this section.  To compute
$\alpha_s$ we used the two loop expression with
$\Lambda_{\mathrm{QCD}}^{(5)} = 0.215$~GeV and $n_{\mathrm{lf}}$
active flavors, where $n_{\mathrm{lf}}$ is the number of flavors with
mass less than the renormalization scale. For the charm quark mass we
used $1.5$ GeV.
The center of mass energy was chosen to be 175 (500) GeV for LEP2
(NLC).  For the renormalization scale we took $\mu =
\sqrt{m^2+(p_t^2(Q)+p_t^2(\overline{Q}))/2}$. In the present process
the choice of scale only affects the value of $\alpha_s$.
We used the
Weizs\"{a}cker-Williams density of \cite{Frix} with an anti-tag angle
$\theta_{\mathrm{max}}$ of 30 (175) mrad for the case of LEP2 (NLC). At the NLC
beamstrahlung is expected to play an important r\^{o}le, so we include its
effect here by adopting for its spectrum the expression given in
\cite{Chen}, with parameters $\Upsilon_{\mathrm{eff}} = 0.039$ and
$\sigma_z=0.5$ mm \cite{Schulte} corresponding to the TESLA design.
For the NLC we will as default coherently superimpose the
Weizs\"{a}cker-Williams density and the beamstrahlung density, in order
to incorporate the case where one photon is of beam- and the other of
bremsstrahlung origin.

For most results we have not used charm-to-D meson fragmentation
function. For the cases that we do, which we indicate explicitly, we
employed the Peterson {\em et al.} parametrization \cite{Peterson}
\begin{equation}
D(z) = \frac{N}{z (1-1/z-\epsilon/(1-z))^2}
\end{equation}
with $\epsilon = 0.06$ the value given in \cite{eps} for the case of
charm. Our interest when including the fragmentation function lies
mainly in how it changes the shapes of distributions, rather than
their normalization. Hence we choose $N$ such that $\int_0^1 dz\,D(z)
= 1$.

We will only present one single particle distribution here, since such
distributions have already been studied in \cite{DKZZ,LPT}.
\begin{figure}[tbhp]

\hspace*{1cm}
\epsfig{file=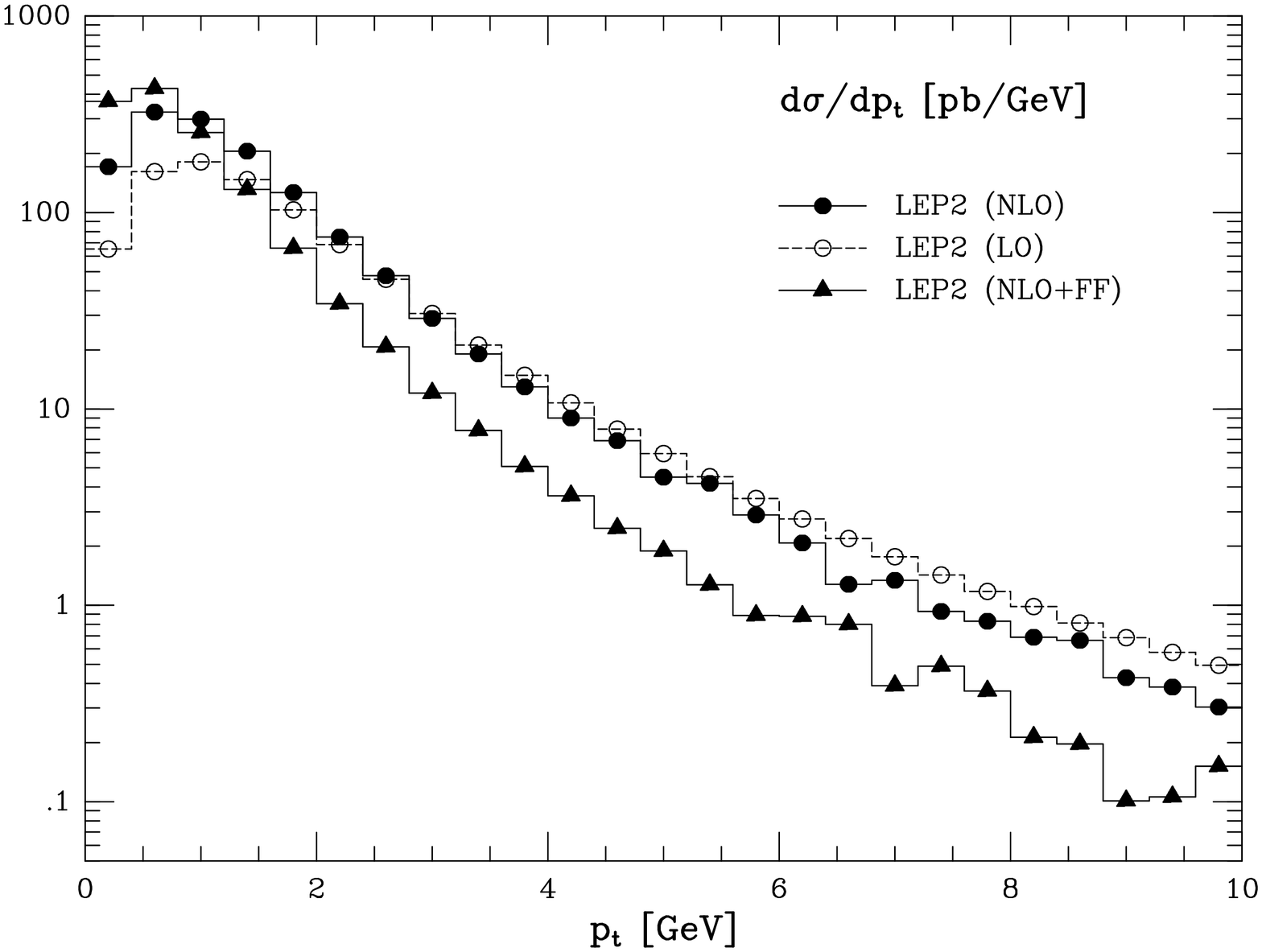,bbllx=0pt,bblly=50pt,bburx=575pt,bbury=800pt,%
        width=10cm,angle=-90}

{\baselineskip=12pt
\noindent {\bf Figure 1:} {\em Single charm quark $p_t$ spectrum at LEP2,
comparison of LO, NLO and NLO with fragmentation.}}

\end{figure}
Fig.1 shows the single particle $p_t$ distribution at LO, NLO, and at
NLO with fragmentation. We see that inclusion of NLO corrections
decreases the cross section at large $p_t$ and enhances it at small
$p_t$, and that the application of the fragmentation function softens
it considerably.

Turning to correlations, we begin by showing
distributions which allow a comparison between the LO and NLO
calculations. In Fig.2 we show the cross section versus the invariant
mass $M_{Q\overline{Q}}$ of the heavy quark pair for LEP2 and NLC at
both LO and NLO.
\begin{figure}[htbp]

\hspace*{1cm}
\epsfig{file=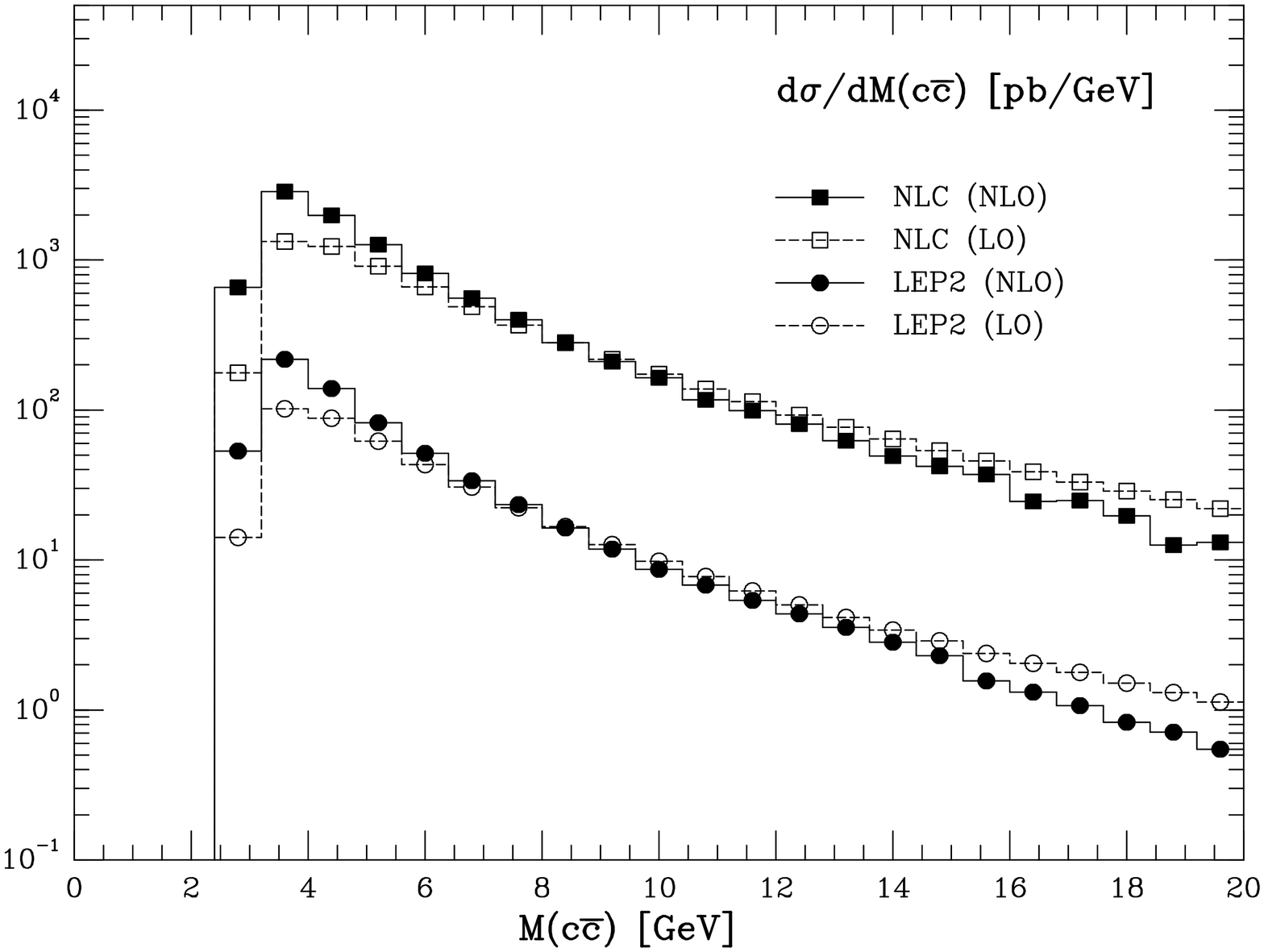,bbllx=0pt,bblly=50pt,bburx=575pt,bbury=800pt,%
        width=10cm,angle=-90}

{\baselineskip=12pt
\noindent {\bf Figure 2:} {\em $M_{Q\overline{Q}}$ distribution for
charm for LEP2 and NLC at both LO and NLO.}}

\end{figure}
Notice in Fig.2 the sizable
difference that occurs at both small and large invariant masses when
including the NLO corrections.  This can be understood as follows.
Consider first the situation where the two photons collide with
all the momentum of their parent leptons.  Denoting the invariant mass
of the heavy quark pair in this case by $\widehat{M}_{Q\overline{Q}}$ ,
then at LO $\widehat{M}_{Q\overline{Q}}$ is fixed at
$\sqrt{s\hphantom{tk}}\!\!\!\!\!_{\gamma\gamma}$. At NLO it may assume
smaller values, and there the cross section is positive.
For $\widehat{M}_{Q\overline{Q}}
= \sqrt{s\hphantom{tk}}\!\!\!\!\!_{\gamma\gamma}$
one has at NLO also a negative contribution coming from the
virtual graphs.  To go back to the case of LEP2 and NLC we must fold
with the photon spectrum. A given $\widehat{M}_{Q\overline{Q}}$ value
then contributes to the spectrum for $M_{Q\overline{Q}}$ under the
restriction $ M_{Q\overline{Q}} < \widehat{M}_{Q\overline{Q}}$, so that
at large $M_{Q\overline{Q}}$ the LO spectrum is mainly modified by the
negative contribution at $\widehat{M}_{Q\overline{Q}}
=\sqrt{s\hphantom{tk}}\!\!\!\!\!_{\gamma\gamma}$, and at small
$M_{Q\overline{Q}}$ by the positive contributions at smaller $
\widehat{M}_{Q\overline{Q}}$.

In Fig.3 we show the $\Delta R$ distribution, defined by $\Delta R =
\sqrt{(\Delta\phi)^2 + (\Delta\eta)^2}$, at LO and NLO for both LEP2
and the NLC.
\begin{figure}[htbp]

\hspace*{1cm}
\epsfig{file=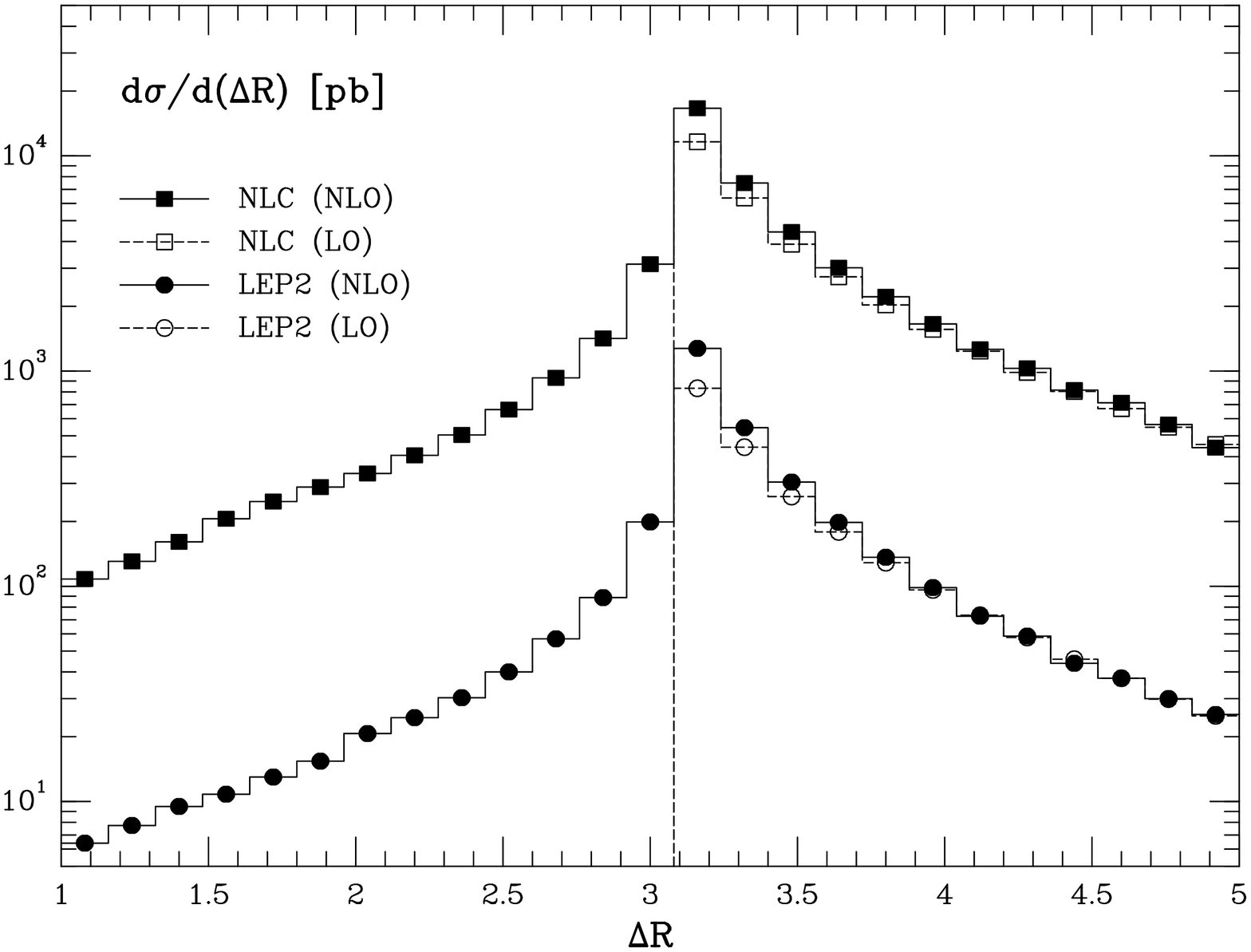,bbllx=0pt,bblly=50pt,bburx=575pt,bbury=800pt,%
        width=10cm,angle=-90}

{\baselineskip=12pt
\noindent {\bf Figure 3:} {\em $\Delta R$ distribution for charm and
anti-charm quark at LEP2 and NLC.}}

\end{figure}
Here $\Delta\phi$ is the azimuthal angle between the charm and
anticharm in the plane transverse to the beam axis and $\Delta\eta$ is
the pseudo-rapidity difference of the two heavy quarks. At LO $\Delta
R>\pi$, but at NLO $\Delta R$ may also assume values below that. Note
that NLO effects seem to be mostly active for $\Delta R\,\,\simlt\,\, 4$.

We now show two distributions which are only non-trivial at NLO (and
higher orders). In Fig.4 we present the $p_t$ distribution of the
charm-anticharm pair, and in Fig.5 the azimuthal correlation between
the two heavy quarks. We also show in Fig.4 the NLC curves with only
beamstrahlung photons and with only Weizs\"{a}cker-Williams photons
for the purpose of comparison.
\begin{figure}[htbp]

\hspace*{1cm}
\epsfig{file=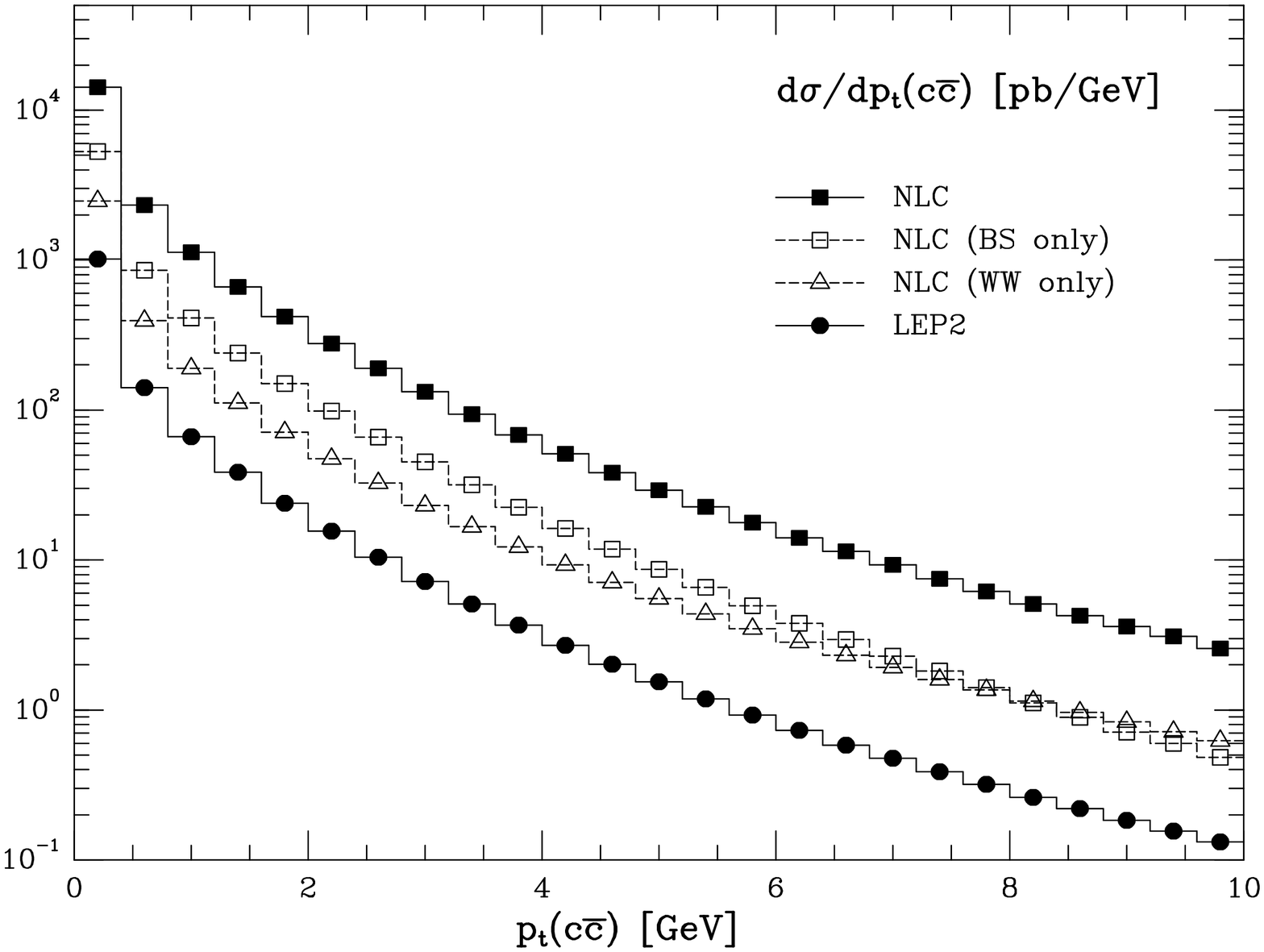,bbllx=0pt,bblly=50pt,bburx=575pt,bbury=800pt,%
        width=10cm,angle=-90}

{\baselineskip=12pt
\noindent {\bf Figure 4:} {\em $p_t(c\bar c)$ distribution for charm
and anti-charm quark at LEP2 and NLC.}}

\end{figure}
\begin{figure}[htbp]

\hspace*{1cm}
\epsfig{file=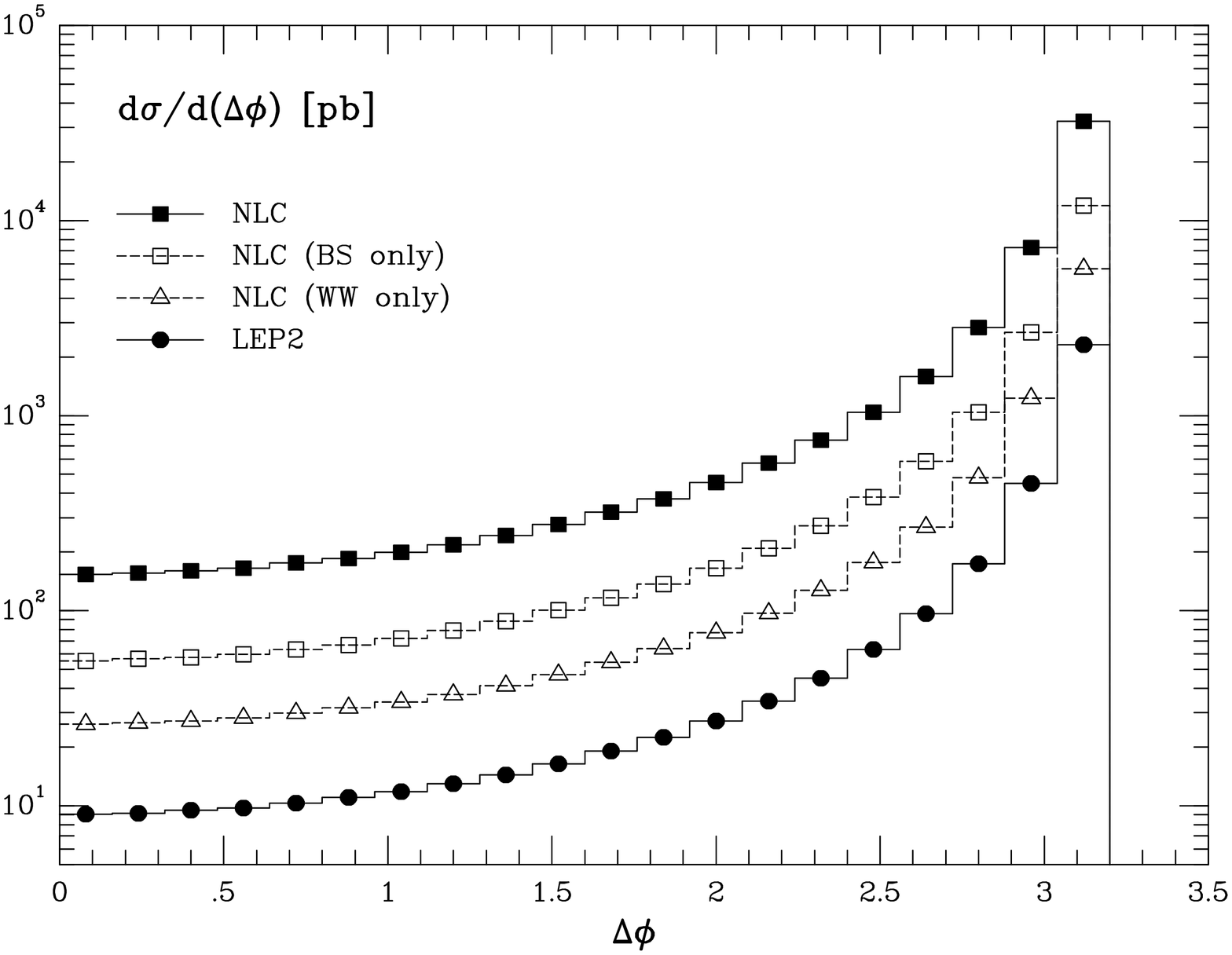,bbllx=0pt,bblly=50pt,bburx=575pt,bbury=800pt,%
        width=10cm,angle=-90}

{\baselineskip=12pt
\noindent {\bf Figure 5:} {\em $\Delta\phi$ distribution for charm and
anti-charm quark at LEP2 and NLC.}}

\end{figure}
One observes in Fig.4 that at the NLC charm pairs produced by
beamstrahlung photons prefer to have a lower $p_t$ than those due to
WW equivalent photons. This is a consequence of the TESLA
beamstrahlung spectrum, which is enhanced at small $z$ and depleted at
large $z$ compared to the WW spectrum ($z$ is the momentum fraction of
the photon relative to its parent lepton).

In Fig.5 we see that the $\Delta\phi$ distributions are all quite
uniform. We observe however that at the NLC for the case of charm the
beamstrahlung contribution dominates the WW one.

Finally we comment on the consequences of choosing different values
of the renormalization scale $\mu$ and the charm mass $m$. To see
how Figs.1-3 change when varying $\mu$ one can simply rescale
the differences between the LO and NLO curves according the change
in $\alpha_s$, whereas in Figs.4 and 5 the whole curve will change
by an overall factor.
We further remark that choosing a different value for $m$ changes
mainly the normalizations of the curves shown in this section,
but not their shapes.

\section{Conclusions}
In this paper we have presented a NLO calculation of heavy quark
production in direct two-photon collisions. We have described our
calculation method and presented numerical studies of various
correlations between the heavy quarks.
We observed that the inclusion of the NLO corrections significantly
modifies the shapes and normalizations of the distributions we studied.
Experimentally such studies
will be challenging at LEP2 due to the low charm acceptance, but they
are certainly feasible at a future $e^+e^-$ linear collider.


\end{document}